\documentclass[%
 reprint,
superscriptaddress,
showpacs,preprintnumbers,
 amsmath,amssymb,
 aps,
prc,
]{revtex4-1}

\usepackage{graphicx}% Include figure files
\usepackage{dcolumn}% Align table columns on decimal point
\usepackage{bm}% bold math
\usepackage{epstopdf}
\usepackage{multirow}
\usepackage{paralist}
\usepackage{booktabs}
\usepackage[version=3]{mhchem}
\usepackage{color}
\usepackage{placeins}
\usepackage{verbatim}
\begin{document}

\preprint{APS/123-QED}

\title{Changes in nuclear structure along the Mn isotopic chain studied via charge radii}
%\title{Structure changes in the Mn isotopes from $N=25$ to $N=39$ studied via charge radii}
%\title{Nuclear structure of the manganese isotopes studied via charge radii systematics}
%\title{Nuclear charge radii of neutron-rich Mn isotopes towards $N=40$}% Force line breaks with \\

\author{H. Heylen}
\email{hanne.heylen@fys.kuleuven.be}
\affiliation{KU Leuven, Instituut voor Kern- en Stralingsfysica, 3001 Leuven, Belgium}

\author{C. Babcock}
\email{cbabcock@triumf.ca}
\affiliation{Oliver Lodge Laboratory, Oxford Street, University of Liverpool, L69 7ZE, United Kingdom}
\affiliation{ISOLDE, Physics Department, CERN, CH-1211 Geneva 23, Switzerland}
\author{R. Beerwerth}
\affiliation{Helmholtz Institute Jena, Fr{\"o}belstieg 3, 07743 Jena, Germany}
\affiliation{Theoretisch-Physikalisches Institut, Friedrich-Schiller-Universit{\"a}t Jena, 07743 Jena, Germany}
\author{J.~Billowes}
\affiliation{School of Physics and Astronomy, The University of Manchester, Manchester M13 9PL, United Kingdom}
\author{M.L.~Bissell}
\affiliation{KU Leuven, Instituut voor Kern- en Stralingsfysica, 3001 Leuven, Belgium}
\affiliation{School of Physics and Astronomy, The University of Manchester, Manchester M13 9PL, United Kingdom}
\author{K.~Blaum}
\affiliation{Max-Planck-Institut f\"ur Kernphysik, D-69117 Heidelberg, Germany}
\author{J.~Bonnard}
\affiliation{Istituto Nazionale di Fisica Nucleare, Sezione di Padova, 35131 Padova, Italy}
\author{P.~Campbell}
\affiliation{School of Physics and Astronomy, The University of Manchester, Manchester M13 9PL, United Kingdom}
\author{B.~Cheal}
\affiliation{Oliver Lodge Laboratory, Oxford Street, University of Liverpool, L69 7ZE, United Kingdom}
\author{T. Day Goodacre}
\affiliation{ISOLDE, Physics Department, CERN, CH-1211 Geneva 23, Switzerland}
\affiliation{School of Physics and Astronomy, The University of Manchester, Manchester M13 9PL, United Kingdom}
\author{D.~Fedorov}
\affiliation{Petersburg Nuclear Physics Institute, 188350 Gatchina, Russia}
\author{S. Fritzsche}
\affiliation{Helmholtz Institute Jena, Fr{\"o}belstieg 3, 07743 Jena, Germany}
\affiliation{Theoretisch-Physikalisches Institut, Friedrich-Schiller-Universit{\"a}t Jena, 07743 Jena, Germany}
\author{R.F.~Garcia Ruiz}
\affiliation{KU Leuven, Instituut voor Kern- en Stralingsfysica, 3001 Leuven, Belgium}
\author{W.~Geithner}
\affiliation{GSI Helmholtzzentrum f\"ur Schwerionenforschung GmbH, D-64291 Darmstadt, Germany}
\author{Ch.~Geppert}
\affiliation{Johannes Gutenberg-Universit\"at Mainz, Institut f\"ur Kernchemie, D-55128, Germany}
\affiliation{Institut f\"ur Kernphysik, TU Darmstadt, D-64289 Darmstadt, Germany}
\author{W.~Gins}
\affiliation{KU Leuven, Instituut voor Kern- en Stralingsfysica, 3001 Leuven, Belgium}
\author{L.K. Grob}
\affiliation{ISOLDE, Physics Department, CERN, CH-1211 Geneva 23, Switzerland}
\affiliation{Institut f\"ur Kernphysik, TU Darmstadt, D-64289 Darmstadt, Germany}
\author{M.~Kowalska}
\affiliation{ISOLDE, Physics Department, CERN, CH-1211 Geneva 23, Switzerland}
\author{K.~Kreim}
\affiliation{Max-Planck-Institut f\"ur Kernphysik, D-69117 Heidelberg, Germany}
\author{S.M.~Lenzi}
\affiliation{Dipartimento di Fisica e Astronomia dell'Universit\`a and INFN, Sezione di Padova, Padova, Italy}
\author{I.D.~Moore}
\affiliation{Department of Physics, University of Jyv\"askyl\"a, PB 35 (YFL) Jyv\"askyl\"a, Finland}
\affiliation{Helsinki Institute of Physics, FI-00014, University of Helsinki, Finland}
\author{B.~Maass}
\affiliation{Institut f\"ur Kernphysik, TU Darmstadt, D-64289 Darmstadt, Germany}
\author{S.~Malbrunot-Ettenauer}
\affiliation{ISOLDE, Physics Department, CERN, CH-1211 Geneva 23, Switzerland}
\author{B.~Marsh}
\affiliation{ISOLDE, Physics Department, CERN, CH-1211 Geneva 23, Switzerland}
\author{R.~Neugart}
\affiliation{Max-Planck-Institut f\"ur Kernphysik, D-69117 Heidelberg, Germany}
\affiliation{Johannes Gutenberg-Universit\"at Mainz, Institut f\"ur Kernchemie, D-55128, Germany}
\author{G.~Neyens}
\affiliation{KU Leuven, Instituut voor Kern- en Stralingsfysica, 3001 Leuven, Belgium}
\author{W.~N\"ortersh\"auser}
\affiliation{Institut f\"ur Kernphysik, TU Darmstadt, D-64289 Darmstadt, Germany}
\author{T.~Otsuka}
\affiliation{Deptarment of Physics, University of Tokyo, Hongo, Bunkyo-ku, Tokyo 113-0033, Japan}
\author{J.~Papuga}
\affiliation{KU Leuven, Instituut voor Kern- en Stralingsfysica, 3001 Leuven, Belgium}
\author{R.~Rossel}
\affiliation{ISOLDE, Physics Department, CERN, CH-1211 Geneva 23, Switzerland}
\author{S.~Rothe}
\affiliation{ISOLDE, Physics Department, CERN, CH-1211 Geneva 23, Switzerland}
\author{R.~S\'anchez}
\affiliation{GSI Helmholtzzentrum f\"ur Schwerionenforschung GmbH, D-64291 Darmstadt, Germany}
\author{Y.~Tsunoda}
\affiliation{Center for Nuclear Study, University of Tokyo, Hongo, Bunkyo-ku, Tokyo 113-0033, Japan}
\author{C.~Wraith}
\affiliation{Oliver Lodge Laboratory, Oxford Street, University of Liverpool, L69 7ZE, United Kingdom}
\author{L.~Xie}
\affiliation{School of Physics and Astronomy, The University of Manchester, Manchester M13 9PL, United Kingdom}
\author{X.F.~Yang}
\affiliation{KU Leuven, Instituut voor Kern- en Stralingsfysica, 3001 Leuven, Belgium}
\author{D.T. Yordanov}
\email{Present address: Institut de Physique Nucl\'eaire, CNRS-IN2P3, Univ. Paris-Sud, Universit\'e Paris-Saclay, 91406 Orsay, France}
\affiliation{Max-Planck-Institut f\"ur Kernphysik, D-69117 Heidelberg, Germany}
%\author{A.P.~Zuker}
%\affiliation{Universit\'e de Strasbourg, IPHC, CNRS, UMR7178, 23 rue du Loess 67037 Strasbourg, France}

\date{\today}

\begin{abstract}
The hyperfine spectra of $^{51,53-64}$Mn were measured in two experimental runs using collinear laser spectroscopy at ISOLDE, CERN. Laser spectroscopy was performed on the atomic \mbox{$3d^5\ 4s^2\ ^{6}\text{S}_{5/2}\rightarrow  3d^5\ 4s4p\  ^{6}\text{P}_{3/2}$} and ionic $3d^5\ 4s\ ^{5}\text{S}_2 \rightarrow 3d^5\ 4p\ ^{5}\text{P}_3$ transitions, yielding two sets of isotope shifts. The mass and field shift factors for both transitions have been calculated in the multiconfiguration Dirac-Fock framework and were combined with a King plot analysis in order to obtain a consistent set of mean-square charge radii which,  together with earlier work on neutron-deficient Mn, allow the study of nuclear structure changes from $N=25$ across $N=28$ up to $N=39$. A clear development of deformation is observed towards $N=40$, confirming the conclusions of the nuclear moments studies. From a Monte Carlo Shell Model study of the shape in the Mn isotopic chain, it is suggested that the observed development of deformation is not only due to an increase in static prolate deformation but also due to shape fluctuations and triaxiality. The changes in mean-square charge radii are well reproduced using the Duflo-Zuker formula except in the case of large deformation.
\end{abstract}

\maketitle

%%% INTRODUCTION
\section{\label{sec:Introduction}Introduction}
The structure of exotic nuclei exhibits many peculiarities such as halo nuclei, islands of inversion  and shape coexistence. Laser spectroscopy has played a decisive role in studying these phenomena as it enables measurements of spins, nuclear moments and changes in mean-square charge radii from the valley of stability to exotic regions of the nuclear chart \cite{Campbell2016}. Due to the complementarity of the measured ground state properties, a comprehensive description of nuclear structure can be established, for example the evolution of both single-particle and collective aspects as a function of nucleon number \cite{Neyens2003,Blaum2013}. Mean-square charge radii in particular have been systematically measured across the nuclear chart. They are sensitive to the shape and size of the nucleus and can also be extracted for even-even nuclei with $I=0$, for which the magnetic moments and spectroscopic quadrupole moments are zero.
 %for all isotopes irrespective of the nuclear spin. % From the theoretical point of view, a wide-scale model which is able to accurately describe mean-square charge radii is unfortunately still lacking, especially when odd-proton chains are considered.

The region between the proton-magic Ca ($Z=20$) and Ni ($Z=28$) isotopes has attracted much attention due to its rich variety in nuclear structure. Around calcium, early measurements revealed a remarkable behaviour of the charge radii between $N = 20$ and $N = 28$ \cite{Touchard1982,Palmer1984,Blaum2008,Avgoulea2011}. Furthermore, the appearance of new magic numbers at $N=32$ and $N=34$ was investigated \cite{Huck1985,Steppenbeck2013,Wienholtz2013,GarciaRuiz2015,GarciaRuiz2016}. Adding a few protons towards the nickel isotopic chain and a few neutrons towards $N = 40$, a rapid development of collectivity is observed, bearing close resemblance to the island of inversion around $N=20$ \cite{Ljungvall2010,Lenzi2010}. 
	\begin{figure}[t!]
	\includegraphics[width=\columnwidth]{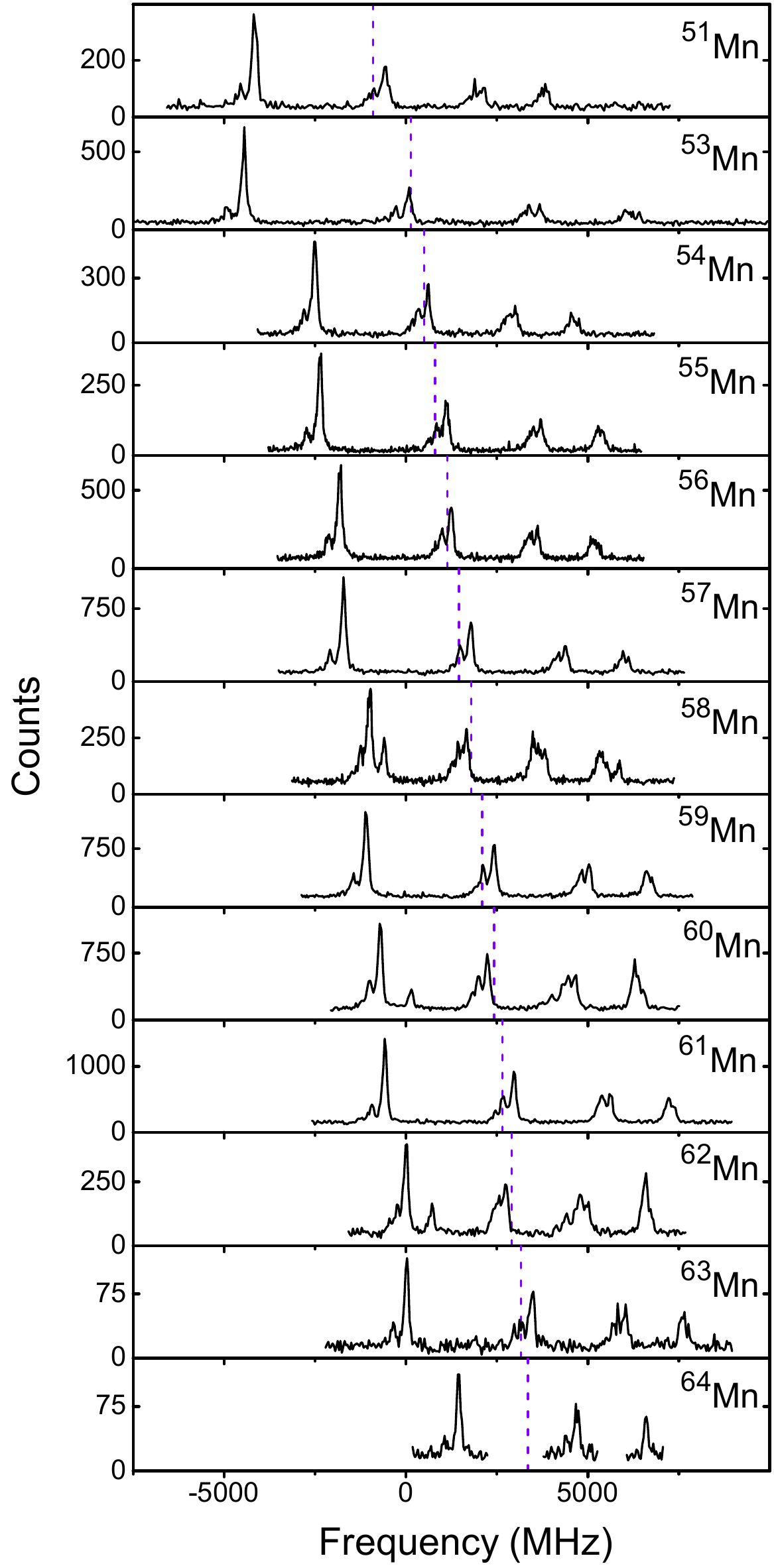}
	\caption{Hyperfine spectra of $^{51,53-64}$Mn measured on the atomic $^{6}\mathrm{S}_{5/2} \rightarrow {}^{6}\mathrm{P}_{3/2}$ transition. The centroid for each spectrum is indicated with a vertical dashed line. For $^{58,60,62}$Mn only the centroid frequency of the ground state is indicated (the isomer shift is too small to be visible on this scale).} 
	\label{Fig:HFS-atom}
	\end{figure}
Although nuclear moments and mean-square charge radii have been essential in the interpretation of the nuclear structure around $N=20$ \cite{Utsuno2004,Himpe2008,Neyens2011,Yordanov2012}, so far the experimental knowledge of these observables around $N=40$ is scarce. Their measurements therefore provide necessary and complementary information.
	\begin{figure}[t!]
	\includegraphics[width=\columnwidth]{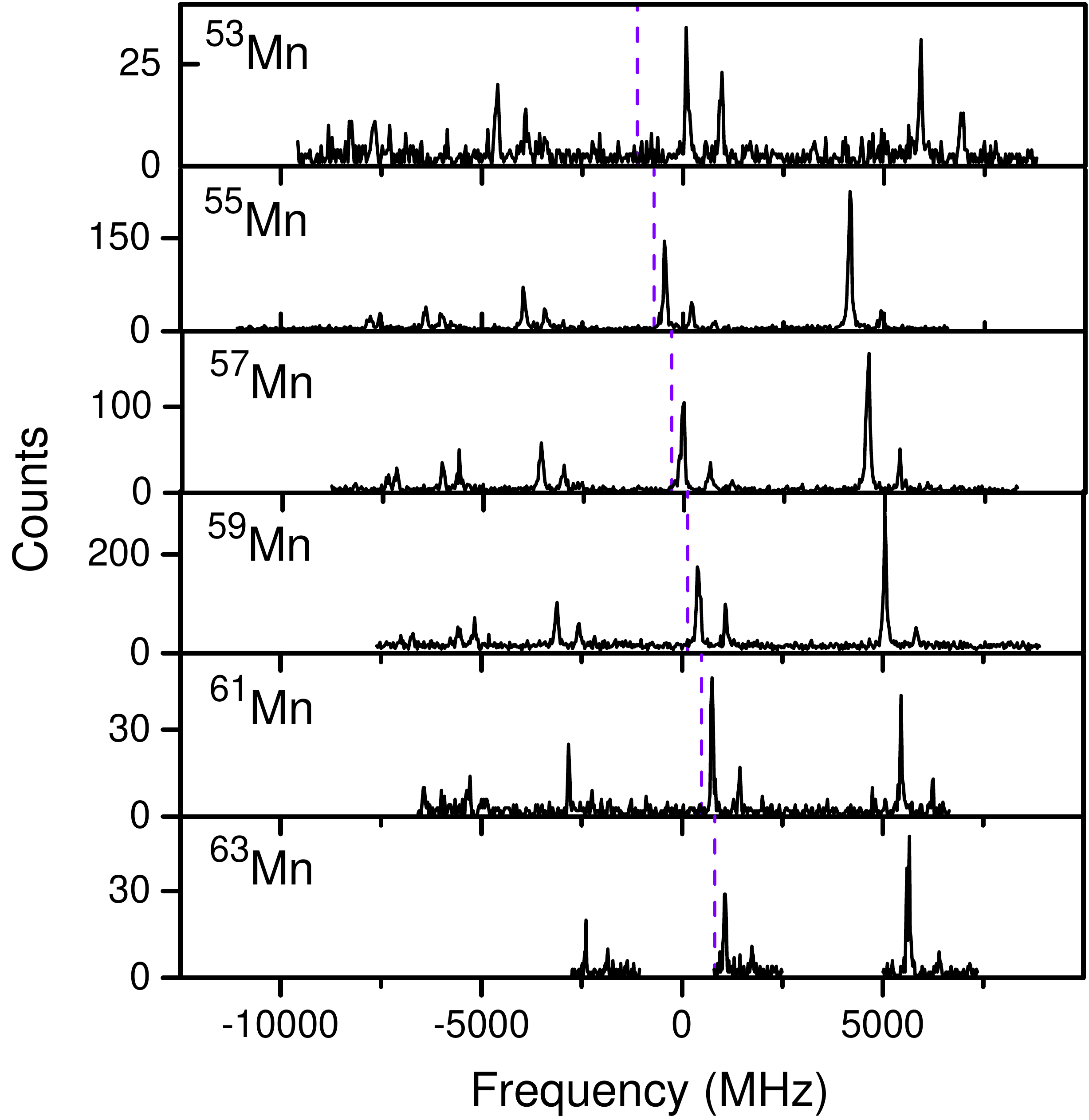}
	\caption{Hyperfine spectra of the odd-even $^{53-63}$Mn isotopes measured on the ionic $^{5}\mathrm{S}_{2} \rightarrow {}^{5}\mathrm{P}_{3}$ transition. The centroid frequency for each spectrum is indicated with a vertical dashed line. Due to the limited yield at $^{63}$Mn only the 3 last multiplets in the hyperfine spectrum were measured, sufficient to extract the centroid and hyperfine parameters. } 
	\label{Fig:HFS-ion}
	\end{figure}
\\ With 25 protons, the manganese isotopes lie centrally between the Ca and Ni isotopic chains. The nuclear moments and mean-square charge radii of Mn between $N=25$ and $N=31$ were previously studied at the IGISOL facility, JYFL, Jyv\"askyl\"a \cite{Charlwood2010}. These measurements have now been extended to $N=39$ in two collinear laser spectroscopy experiments at ISOLDE, CERN \cite{Babcock2015,Heylen2015,Babcock2016}. With the ground state properties of Mn known from $N=25$ across $N=28$ up to $N=39$, a detailed picture of the changing nuclear structure emerges. The mean-square charge radii and quadrupole moments show a clear kink at $N=28$ associated with a neutron shell closure, while the effect of $N=28$ on the two-neutron separation energies is not immediately apparent \cite{Charlwood2010}. The steeply increasing slope of the radii between $N=28$ and $N=32$ is found to be  common to  the isotopic chains in this region, indicating an almost $Z$-independent core-polarization by neutrons filling the upper $pf$-shell \cite{Kreim2014}. Going even more neutron-rich, the Mn magnetic and quadrupole moments show evidence of increased collectivity towards $N=40$ \cite{Babcock2015,Heylen2015,Babcock2016}. Comparisons with large-scale shell model calculations have illustrated the importance of particle-hole excitations across $N=40$ and  $Z=28$ for the description of the nuclear structure approaching $N=40$.
\\ In this article, the mean-square charge radii of $^{51,53-64}$Mn measured in the two previously mentioned ISOLDE experiments are presented. Mn is to date the first isotopic chain below $Z=28$ for which the ground state properties beyond $N=28$ can be systematically studied for such a long sequence of isotopes.

%%% EXPERIMENTAL METHOD
\section{Experimental method}
%\begin{table*}[t!] % for pagewide tables

The experimental campaign was performed at the ISOLDE radioactive ion beam facility in CERN where exotic manganese isotopes were produced via the bombardment of a thick uranium carbide target with 1.4-GeV protons. Element selective ionization was achieved using the resonance ionization laser ion source (RILIS) \cite{Rothe2016}. After mass separation in the HRS, the isotopically pure ion beam was cooled and bunched in the gas-filled RFQ ISCOOL \cite{ISCOOL} and reaccelarated to, respectively, 40 keV and 30 keV in the first and second experiment. The bunched beam was then guided to the dedicated collinear laser spectroscopy beam line COLLAPS \cite{Mueller1983,Nortershauser2010,Papuga2014} where it was overlapped with a co-propagating narrowband laser beam. A Doppler-tuning potential was applied to scan the laser frequency across the hyperfine structure in the reference frame of the ions/atoms. The fluorescence light from the decay following the resonant laser-induced excitations between the hyperfine levels was detected using sensitive photomultiplier tubes placed perpendicular to the beam line. 
\\ In a first experiment, laser spectroscopy was performed on an atomic transition  from the $3d^5\ 4s^2\ ^{6}\text{S}_{5/2}$ ground state to the $3d^5\ 4s4p\  ^{6}\text{P}_{3/2}$ excited state at a wavelength of 280.1907 nm (in vacuum). Hyperfine spectra of $^{51,53-64}$Mn were obtained. The details of the experiment and data analysis are described in \cite{Babcock2015, Heylen2015}. To improve the sensitivity to quadrupole moments, a second experiment was performed on an ionic transition at 295.0066 nm (in vacuum) from the $3d^5\ 4s\ ^{5}\text{S}_2$ metastable state at 9472.97 cm$^{-1}$ to the $3d^5\ 4p\ ^{5}\text{P}_3$ state at 43370.51 cm$^{-1}$. The population of the ionic metastable state was enhanced using the technique of optical pumping in the ISCOOL cooler/buncher, as outlined in \cite{Babcock2016}. In this second experiment, only hyperfine spectra of the odd-even $^{53-63}$Mn isotopes were measured.

%%% RESULTS
\section{Results}
	
Typical hyperfine spectra measured on the atomic  and ionic transition are shown in  Fig.~\ref{Fig:HFS-atom} and \ref{Fig:HFS-ion}, respectively. The centroid frequency $\nu^A_0$ for each spectrum is indicated with a vertical dashed line. The isotope shifts relative to $^{55}$Mn 
$$\delta \nu^{55,A} = \nu^A_{0}- \nu^{55}_{0}$$
 extracted from these hyperfine spectra are presented in Table \ref{Table:IS}, the hyperfine $A$ and $B$ parameters were reported earlier in \cite{Babcock2015, Heylen2015, Babcock2016}. Statistical errors obtained in the fit procedure are indicated with round brackets while a systematic error due to a $\pm$15 V uncertainty in the acceleration voltage calibration is assumed for both experiments, indicated in square brackets. For completeness, also the isotope shifts measured previously at the IGISOL facility JYFL in Jyv\"askyl\"a \cite{Charlwood2010} using the same ionic transition are reported in Table \ref{Table:IS}.
 \\ The measured isotope shifts can be related to the changes in mean-square charge radii $\delta \langle r^2\rangle$ using (see e.g.\ \cite{Blaum2013})
	\begin{equation}
	\delta \nu^{55,A}_j = M_j \frac{m_{A}-m_{55}}{m_{A}m_{55}} + F_j \delta \langle r^2\rangle^{55,A}.
	\label{chargeradii-extraction}
	\end{equation}
Here $m_A$ is the mass of an isotope $A$ and $M_j$ and $F_j$ are the mass and field shift parameters specific to the electronic transition $j$. The first term (mass shift) is due to the recoil kinetic energy of a nucleus with finite mass while the second term (field shift) originates from the changes in nuclear charge distribution and hence contains the charge radius dependence. To reliably extract the changes in mean-square charge radii, an accurate knowledge of both electronic parameters is required. In first-order perturbation theory, these parameters are constant along an isotopic chain but they need to be determined for each electronic transition separately.
\begin{table}[t]  % for columnwide tables
\caption{Mn isotope shifts relative to $^{55}$Mn measured on the atomic \mbox{$3d^5\ 4s^2\ ^{6}\mathrm{S}_{5/2}\rightarrow  3d^5\ 4s4p\  ^{6}\mathrm{P}_{3/2}$} transition and the ionic $3d^5\ 4s\ ^{5}\mathrm{S}_2 \rightarrow 3d^5\ 4p\ ^{5}\mathrm{P}_3$ transition. Additional to the statistical uncertainty in round brackets, the systematic uncertainty on the isotope shifts due to a $\pm$15V uncertainty on the calibration of the acceleration voltage is indicated in square brackets.}
\begin{ruledtabular}
\begin{tabular}{l c r r r r l l l l l l l }
\\ [-2.3ex]
$A$ & $N$ & $\delta \nu ^{55,A}_\text{atom}$ (MHz) &  $\delta \nu^{55,A}_\text{ion}$ (MHz) \\
 [0.45ex]
\colrule \\ [-1.85ex]
$50g$ & 25 & \multicolumn{1}{c}{-} & $-$1573(2)[27]\footnotemark[1] \\
$50m$ & 25 &\multicolumn{1}{c}{-} &  $-$1514(8)[27]\footnotemark[1] \\
$51$ & $26$ &  $-$1698(7)[20] &  $-$1201(26)[21]\footnotemark[1]  \\
$52g$ & 27 & \multicolumn{1}{c}{-}& $-$745(7)[16]\footnotemark[1] \\
$52m$ & 27 & \multicolumn{1}{c}{-} &  $-$782(3)[16]\footnotemark[1] \\
$53$ & $28$ &  $-$669(6)[9] &  $-$418(3)[10] \\
 & & &  $-$418(2)[10]\footnotemark[1] \\
$54$ & $29$ & $-$307(6)[5] &  $-$192(4)[5]\footnotemark[1]  \\
$55$ & $30$ & 0  & 0  \\
$56$ & $31$ & 340(7)[5] & 240(6)[5]\footnotemark[1] \\
$57$ & $32$ & 655(7)[9] & 422(5)[10] \\
$58g$ & $33$ & 983(4)[13] & \multicolumn{1}{c}{-}  \\
$58m$ & 33 & 990(4)[13]  &  \multicolumn{1}{c}{-} \\
$59$ & $34$ & 1284(5)[17]  & 863(11)[19]\\
$60g$ & $35$ & 1607(10)[21]  &\multicolumn{1}{c}{-} \\
$60m$ & 35  & 1594(4)[21] &\multicolumn{1}{c}{-} \\
$61$ & $36$ & 1843(7)[25] & 1215(2)[27]  \\
$62g$ & $37$ & 2099(8)[28] & \multicolumn{1}{c}{-}  \\
$62m$ & 37  & 2088(5)[28] & \multicolumn{1}{c}{-} \\
$63$ & $38$ & 2355(6)[32] &  1542(5)[35]  \\
$64$ & $39$ & 2566(6)[35] &  \multicolumn{1}{c}{-}\\
\end{tabular}
\end{ruledtabular}
\footnotetext[1]{Isotope shift measured in \cite{Charlwood2010} including a systematic error arising from a $\pm$15V uncertainty of the calibration in the acceleration voltage.}
\label{Table:IS}
\end{table}
\subsection{Electronic factor calculations}
	\begin{figure}[t]
	\includegraphics[width=\columnwidth]{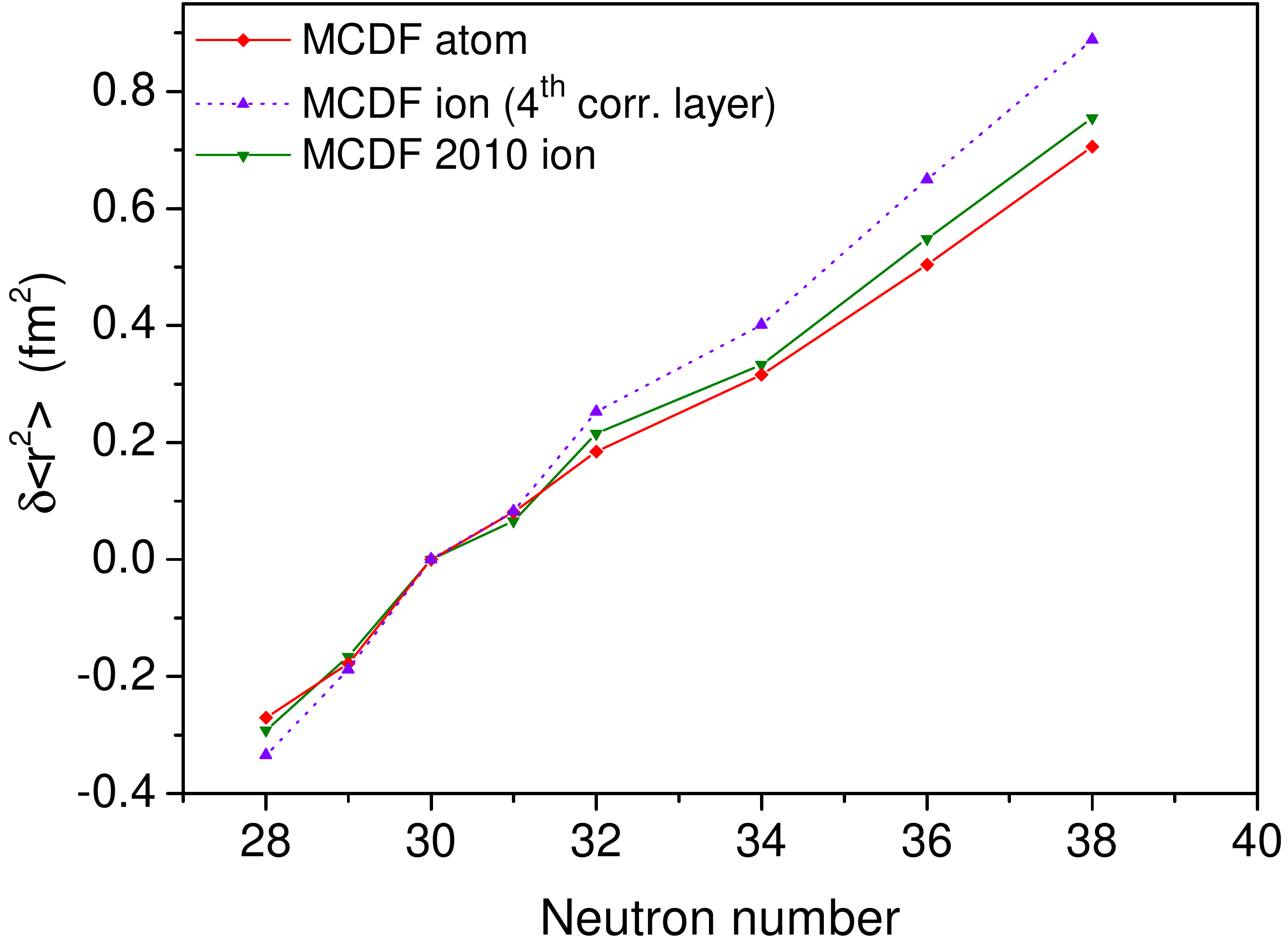}
	\caption{(Color online) Comparison between the changes in mean-square charge radii between $N=28$ and $N=39$ calculated using the mass and field shift factors obtained in three different MCDF calculations. Only the statistical experimental uncertainties are shown.}
	\label{Fig:MCDF-comparison}
	\end{figure}
	
\label{subsection:electronicfactors}
For Mn, the mass and field shift factors cannot be determined by a direct comparison between the measured isotope shifts and the independently known charge radii since only the absolute charge radius of one isotope is known from non-optical methods. Therefore, one has to rely on theoretical calculations to obtain the electronic $M$ and $F$ factors, using e.g.\ the multiconfiguration Dirac-Fock (MCDF) method \cite{Cheal2012}.
\\ The MCDF method, implemented in the {\sc Grasp2k} code \cite{Jonsson2013}, is applied to generate the orbitals for the atomic $3d^5 4s^2 \ {}^6\text{S}_{5/2}$ and $3d^5 4s 4p\ {}^6\text{P}_{3/2}$ configurations as well as the ionic $3d^5 4s\ {}^5\text{S}_2$ and $3d^5 4p\ {}^5\text{P}_3$ levels. In a subsequent step, the Breit interaction is included by first-order perturbation theory in the configuration-interaction method \cite{Fritzsche2002b}. The resulting wave functions are then used to compute the mass shift parameters by utilizing the {\sc ris3} module \cite{Naze2013} and the field shift parameter is computed using the {\sc Ratip} tools \cite{Fritzsche2012}. This distinguishes the current results from the previous computations on the same ionic transition reported in \cite{Charlwood2010}, where the relativistic correction to the mass shift operator was neglected. However, these corrections do not exceed the uncertainty due to the complex electronic structure. The open $3d$ shell makes theoretical calculations on atomic and singly ionized manganese very challenging and extensive in terms of required computational power.
	\begin{figure}[t]
	\includegraphics[width=\columnwidth]{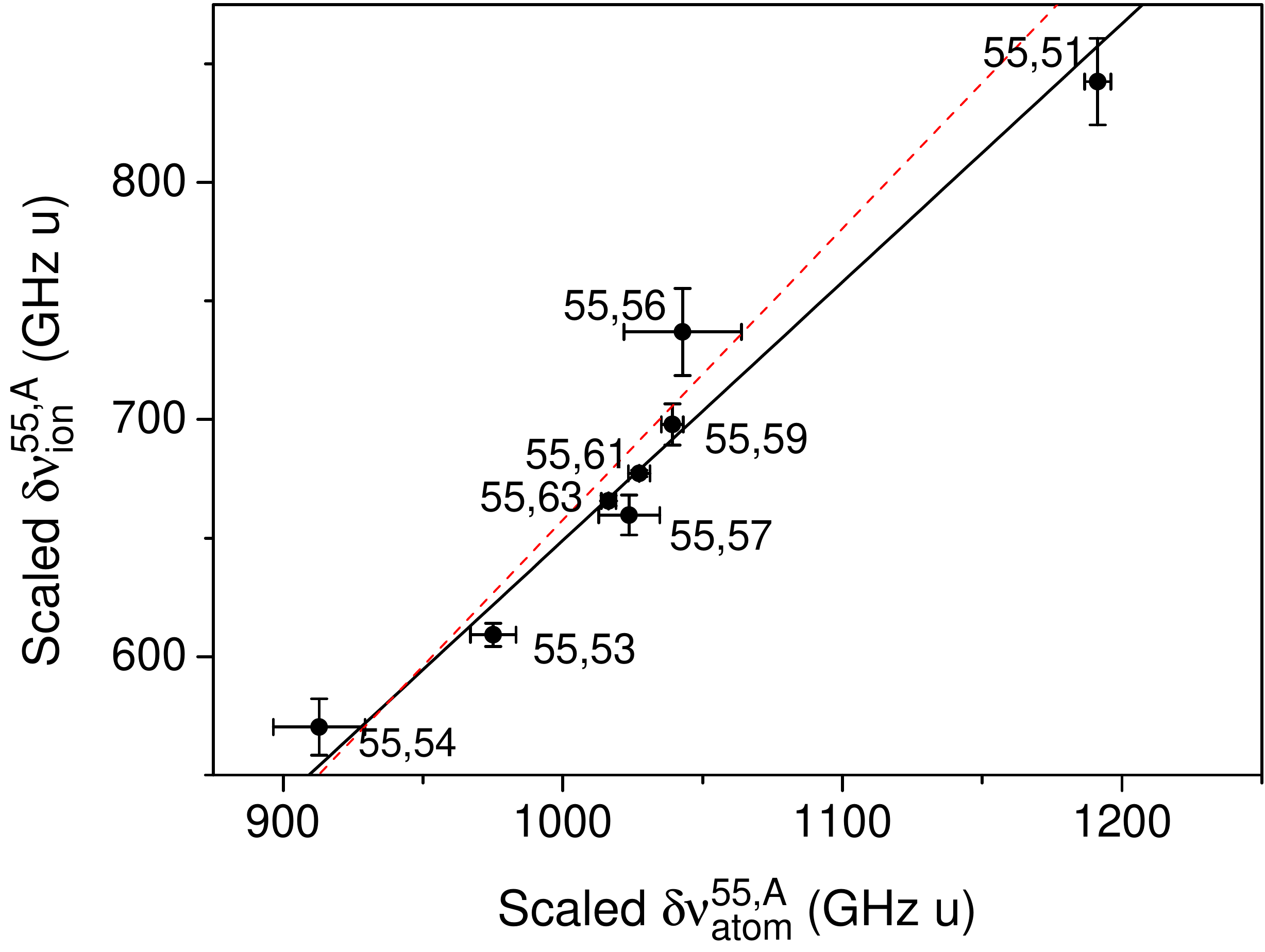}
	\caption{(Color online) King plot using the isotope shifts of the atomic and ionic transitions, shown in Table~\ref{Table:IS}. The best linear fit (black solid line) is shown together with the linear relation obtained by filling in the atomic MCDF and the ionic MCDF 2010 factors in eq.\ (\ref{eq:KP}) (dashed red line). }
	\label{Fig:Kingplot}
	\end{figure}
\begin{table} % for columnwide tables
\caption{Overview of the electronic mass and field shift factors for the atomic \mbox{$3d^5\ 4s^2\ ^{6}\text{S}_{5/2}\rightarrow  3d^5\ 4s4p\  ^{6}\text{P}_{3/2}$}  transition and the ionic $3d^5\ 4s\ ^{5}\text{S}_2 \rightarrow 3d^5\ 4p\ ^{5}\text{P}_3$ transition calculated in the multiconfiguration Dirac-Fock framework. Also the atomic factors calculated with respect to (wrt) the ionic factors using the King plot analysis are shown and vice versa. The estimated theoretical uncertainties on the parameters are indicated in square brackets, and where applicable, the fit errors from the King plot are indicated in round brackets. For details about the empirically modified parameters, see section \ref{sec:Discussion}.}
\begin{ruledtabular}
\begin{tabular}{c l l l }
\\ [-2.3ex]
Transition & Method &  $M$ (GHz u) & $F$ (MHz/fm$^2$) \\
 [0.45ex]
\colrule \\ [-1.85ex]
Atom:  & \textbf{MCDF} & \textbf{+1158}[81] &  $\boldsymbol{-465}[70]$   \\
 $^{6}\text{S}_{5/2} \rightarrow {}^{6}\text{P}_{3/2}$ & King plot wrt & \multirow{2}{*}{+1188(108)[65]} & \multirow{2}{*}{$-525$(39)[93]} \\
&  \ MCDF 2010 \\
 & Emp. Mod. & +1215 & $-465$ 
 \\ [2ex]
Ion: & MCDF 4$^\text{th}$ & +875[131]  & $-546[82]$  \\
$^{5}\text{S}_{2} \rightarrow {}^{5}\text{P}_{3}$ & MCDF 2010 \cite{Charlwood2010} & +852[60]   & $-572[86]$  \\
 &\textbf{King plot wrt} &\multirow{2}{*}{ \textbf{+819}(125)[88]} & \multirow{2}{*}{$\boldsymbol{-506}(38)[76]$} \\ 
 &\textbf{ MCDF atom} & & \\
& Emp. Mod. & +881(129) & $-506(38)$ \\
\end{tabular}
\end{ruledtabular}
\label{Table:electronicparameters}
\end{table}
\begin{table*}[tp] % for pagewide tables
%\begin{table}[t!]  % for columnwide tables
\caption{Changes in mean-square charge radii of $^{50-64}$Mn relative to $^{55}$Mn extracted from the isotope shifts measured on the atomic $^{6}\text{S}_{5/2} \rightarrow {}^{6}\text{P}_{3/2}$ and the ionic $^{5}\text{S}_{2} \rightarrow {}^{5}\text{P}_{3}$ transition. On the lefthand side of the table, the charge radii are obtained with the electronic parameters shown in bold in Table \ref{Table:electronicparameters}. On the righthand side of the table, the charge radii are obtained with the empirically modified electronic parameters, as explained in section \ref{sec:systematics}. The statistical uncertainties and the systematic uncertainties due to the uncertainty on the acceleration voltage are shown in round and square brackets, respectively. No theoretical uncertainties due to the uncertainty on the calculated mass and field shift factors are shown, as explained in section \ref{sec:results-chargeradii}. }
\begin{ruledtabular}
\begin{tabular}{c c c| r r c |r r l l l l l l l l }
\\ [-2.3ex]
$A$ & $N$ & &\multicolumn{2}{c}{ $\delta \langle r^2 \rangle^{55,A}$ (fm$^2$)} & & \multicolumn{2}{c}{ $\delta \langle r^2 \rangle^{55,A}_\text{emp. mod.}$ (fm$^2$)} \\
 & & & \multicolumn{1}{c}{atom} & \multicolumn{1}{c}{ion} & & \multicolumn{1}{c}{atom} &\multicolumn{1}{c}{ion} \\
 [0.45ex]
\colrule \\ [-1.85ex]
$50g$ & 25 & & \multicolumn{1}{c}{-}  & 0.168(4)[53]& &\multicolumn{1}{c}{-} &  $-$0.055(4)\\
$50m$ &  25 &  &\multicolumn{1}{c}{-}  & 0.051(16)[53]& & \multicolumn{1}{c}{-}  &  $-$0.171(16) \\
$51$ & $26$ & & 0.102(14)[42]& 0.065(53)[41]&  & $-$0.073(14)&  $-$0.11(5)\\
$52g$ & 27 & &\multicolumn{1}{c}{-}  &  $-$0.226(14)[31]&  &\multicolumn{1}{c}{-}  &  $-$0.354(14)\\
$52m$ & 27 & &\multicolumn{1}{c}{-} &  $-$0.153(6)[31] &  &\multicolumn{1}{c}{-}  &  $-$0.281(6)\\
$53$ & $28$ &  & $-$0.270(12)[20]&  $-$0.285(7)[20]& & $-$0.354(12)  & $-$0.369(7)\\
$54$ & $29$ &  & $-$0.177(12)[10]&  $-$0.166(8)[10]&  &$-$0.219(12)&  $-$0.207(8) \\
$55$ & $30$ & & \multicolumn{1}{c}{0}  & \multicolumn{1}{c}{0}& & \multicolumn{1}{c}{0}& \multicolumn{1}{c}{0} \\
$56$ & $31$ & & 0.081(15)[10]& 0.053(12)[10]& &0.121(15)& 0.093(12)\\
$57$ & $32$ & & 0.185(15)[19] & 0.202(11)[19]& &0.263(15)& 0.280(11)  \\
$58g$ & $33$ & &0.234(8)[28]& \multicolumn{1}{c}{-}  & &0.350(8) & \multicolumn{1}{c}{-}   \\
$58m$ & 33 &  & 0.220(8)[28]&  \multicolumn{1}{c}{-}  & &0.336(8) & \multicolumn{1}{c}{-}\\
$59$ & $34$ &  & 0.316(10)[37]  &  0.297(21)[37]& & 0.467(10) &0.448(21)\\
$60g$ & $35$ & & 0.329(14)[45] &\multicolumn{1}{c}{-}  & & 0.515(14) & \multicolumn{1}{c}{-} \\
$60m$ & 35  &   &0.357(9)[45]& \multicolumn{1}{c}{-}  & & 0.544(9) & \multicolumn{1}{c}{-} \\
$61$ & $36$ &   &0.504(15)[53] & 0.504(5)[54] & & 0.724(15) &0.724(5)\\
$62g$ & $37$ &  & 0.615(18)[61] & \multicolumn{1}{c}{-}   & & 0.868(18) &\multicolumn{1}{c}{-} \\
$62m$ & 37  &  & 0.640(9)[61] & \multicolumn{1}{c}{-}   & & 0.893(9) &\multicolumn{1}{c}{-}\\
$63$ & $38$ &   &0.706(13)[69] &  0.704(10)[69]& & 0.990(13) &0.988(10)  \\
$64$ & $39$ &  & 0.873(14)[76] &  \multicolumn{1}{c}{-}  & & 1.188(14) &\multicolumn{1}{c}{-} 
\end{tabular}
\end{ruledtabular}
\label{Table:chargeradii}
\end{table*}
\\ In the MCDF approach, the wave function is represented as a superposition of configuration state functions (CSFs). Each CSF represents a specific electronic configuration and the choice of this basis set is crucial for the quality of the obtained results. In the lowest-order approximation, the model space consists of the $3d$, $4s$ and $4p$ valence shells. This space is then extended by adding a set of correlation orbitals $nl$, which are populated by virtual excitations from the reference configurations. Due to computational restrictions, the correlation orbitals are limited to orbitals with angular momentum up to $f$. In subsequent steps, the model space is systematically extended by further layers with a new principal quantum number $n$. The virtue of this model is that it allows convergence of the desired quantities to be monitored as the model space is expanded. For all computations performed in this work, single and double excitations from the valence shells, as well as single excitations from all core shells, are taken into account. All calculations are performed with common core orbitals, while the correlation orbitals are separately optimized for the upper and lower level.
\\ For the atomic calculations, the reference configurations $3d^5 4(s^2 + p^2)$ for the ground state and the excited state configurations $3d^5 4s 4p$ and $3d^6 4s$ are used to generate five correlation layers. This choice of the reference configurations yields a well-balanced basis-expansion such that the transition energy as well as the $M$ and $F$ parameters converge well. The main source of error comes from omitted electronic correlations. Because these contributions are difficult to quantify, a conservative theoretical uncertainty of 15\% on $M$ and $F$ is assumed.
\\ The computations on the ionic transitions show much more complex electron correlations. It is found that the two levels of interest significantly correlate with configurations where at least one electron is excited to the $4f$ or $4d$ shell. A complete treatment of this kind of correlations is not possible due to the extremely large basis expansions. Therefore, an alternative approach is used where the zeroth-order approximation is computed with a space spanned by the $3d^5 4s$, $3d^4 4p^2$ and $3d^3 4s 4d^2$ configurations for the lower level and $3d^5 4p$, $3d^4 4s 4p$ and $3d^3 4p 4d^2$ for the upper level. The  correlation layers were then generated by using only the $3d^5 (4s + 4p)$ configurations as a reference. This approach yields good convergence for the first four layers, however, convergence fails for the fifth layer, possibly due to neglected electronic correlations.
\\ An overview of the different electronic factors obtained from the computations described above and in \cite{Charlwood2010} (labeled MCDF 2010) can be found in Table~\ref{Table:electronicparameters} and a comparison of  the changes in charge radii evaluated with those factors is shown in 
 Fig.~\ref{Fig:MCDF-comparison}. There is good agreement between the charge radii obtained with the electronic factors for the atomic transition calculated in this work and the previously published calculations on the ionic transition. The computation on the ionic transition including up to four correlation layers (this work) yields slightly larger differences in charge radii, but the agreement is still better than the above estimated uncertainty. Due to the aforementioned convergence problems of the ionic calculations and the limited treatment of electronic correlations, the atomic calculations are considered to be more reliable. Therefore, the atomic results are used in the following King plot analysis.
\\ An additional consistency check of the calculations is performed via the King plot method \cite{King}
\begin{equation}
\small
\mu^{55,A}\delta \nu^{55,A}_\text{ion} = \frac{F_\text{ion}}{F_\text{atom}}\mu^{55,A}\delta \nu^{55,A}_\text{atom} + M_\text{ion}-\frac{F_\text{ion}}{F_\text{atom}}M_\text{atom} 
\label{eq:KP}
\end{equation}
\normalsize
where $\mu^{55,A} = m^{55}m^{A}/(m^{A}-m^{55})$ is used as a scaling factor for the isotope shifts. In Fig.~\ref{Fig:Kingplot} the scaled isotope shifts of the ionic transition are plotted against those of the atomic transition. A linear fit then gives the slope $ \frac{F_\text{ion}}{F_\text{atom}}  = 1.09(8)$ and intercept \mbox{$M_\text{ion}-\frac{F_\text{ion}}{F_\text{atom}}M_\text{atom}  = -44(8)\ \cdot10^4 \text{ MHz u}$} from which the mass and field shift parameters of the atomic transition can be determined from the ionic computation and vice versa (see Table \ref{Table:electronicparameters}). 
\\Comparison with the independently computed values then permits the estimation of the uncertainty of the computations. This suggests that the above quoted uncertainty of 15\% on $F$ is realistic while a 15\% uncertainty on $M$ seems overestimated. From the comparison to regional systematics in section \ref{sec:Discussion} and the agreement of the King plot analysis, an uncertainty of 7\% on $M$ seems to be more appropriate.

\subsection{Changes in mean-square charge radii}	\label{sec:results-chargeradii}

The changes in mean-square charge radii relative to $^{55}$Mn are determined using the experimental isotope shifts in Table~\ref{Table:IS} and the electronic factors shown in bold in Table~\ref{Table:electronicparameters}. Because a King plot is used to link the mass and field shift factors of the ionic transition to the atomic transition, the charge radii extracted from the two transitions are consistent with each other, as can be seen in Table~\ref{Table:chargeradii}. Nevertheless, since the charge radii are obtained by using only two parameters for each transition, the one-by-one agreement between the two transitions provides an independent cross check of the consistency of the isotope shift measurements. 
\\ In Table~\ref{Table:chargeradii} no theoretical uncertainties are shown. This is justified because although the theoretical uncertainty on the electronic factors gives a large uncertainty on the exact $\delta \langle r^2 \rangle$ value, it does not alter the discussion in section \ref{sec:Discussion}. Indeed, a variation of $M$ and $F$ causes a general pivot about the reference point but does not change the local trends. Since nuclear structure effects manifest themselves as local irregularities in the course of the charge radii, the interpretation is not affected by the theoretical uncertainty.
\\ Details on the empirically modified charge radii shown on the righthand side of Table~\ref{Table:chargeradii} will be given in the next section. Note that in this case only statistical uncertainties are shown because the systematic uncertainty on the isotope shifts is incorporated effectively in the modified mass shift parameter \cite{Marinova2011}.

%%% DISCUSSION
\section{\label{sec:Discussion}Discussion}
\begin{figure}[t]
	\includegraphics[width=0.95\columnwidth]{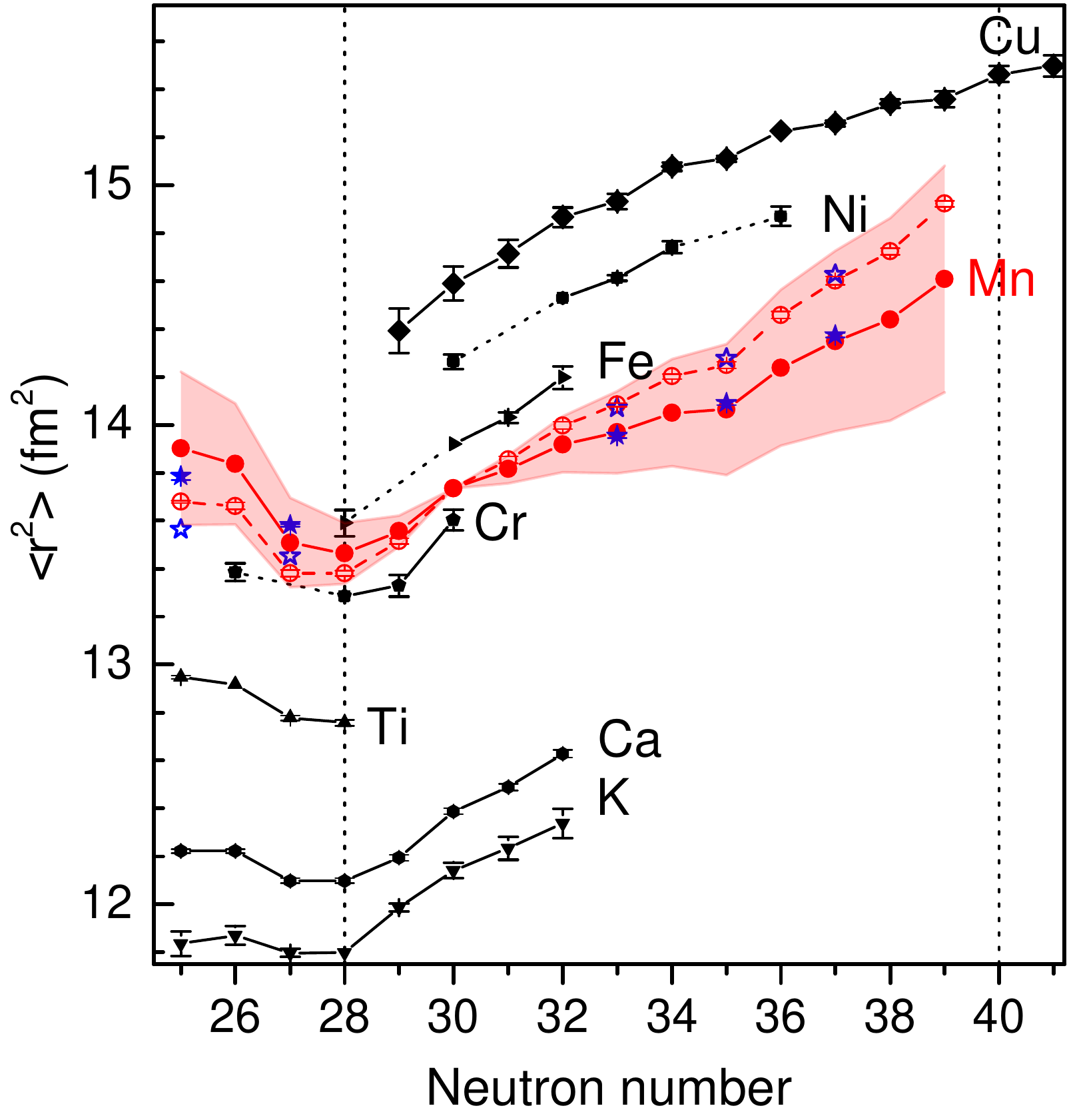}
	\caption{(Color online) The experimental mean-square charge radii known in the Mn region between $N= 25$ and \mbox{$N=39$}. The Mn isotopes are highlighted in red with the isomers in $^{50,52,58,60,64}$Mn indicated by a blue star. The Mn charge radii obtained using the MCDF atomic parameters and King plot ionic parameters are indicated by full symbols connected by a solid line. The shaded area represents the systematic uncertainty assuming 15\% uncertainty on $F$ and 7\% on $M$. The empirically modified Mn charge radii are shown by open symbols connected with a dashed line (see text).}
\label{Fig:chargeradii-systematics}
\end{figure}

\subsection{Mn radii systematics}\label{sec:systematics}	
In Fig.~\ref{Fig:chargeradii-systematics} the mean-square charge radii $\langle r^2 \rangle$ of the $^{50-64}$Mn isotopes ($N=25-39$) are plotted along with the charge radii of the neighboring isotopes between $_{19}$K and $_{29}$Cu. For Mn the results from the atomic transition are shown except for $^{50,51,52,54,56}$Mn for which only ionic data is available. The $\langle r^2 \rangle$ values are obtained with the reference radii of Fricke and Heilig \cite{FrickeHeilig} and the published changes in mean-square charge radii \cite{Kreim2014,GarciaRuiz2016,Gangrsky2004,Benton1997,Steudel1980,Bissell2016}.
\begin{figure}[t]
	\includegraphics[width=\columnwidth]{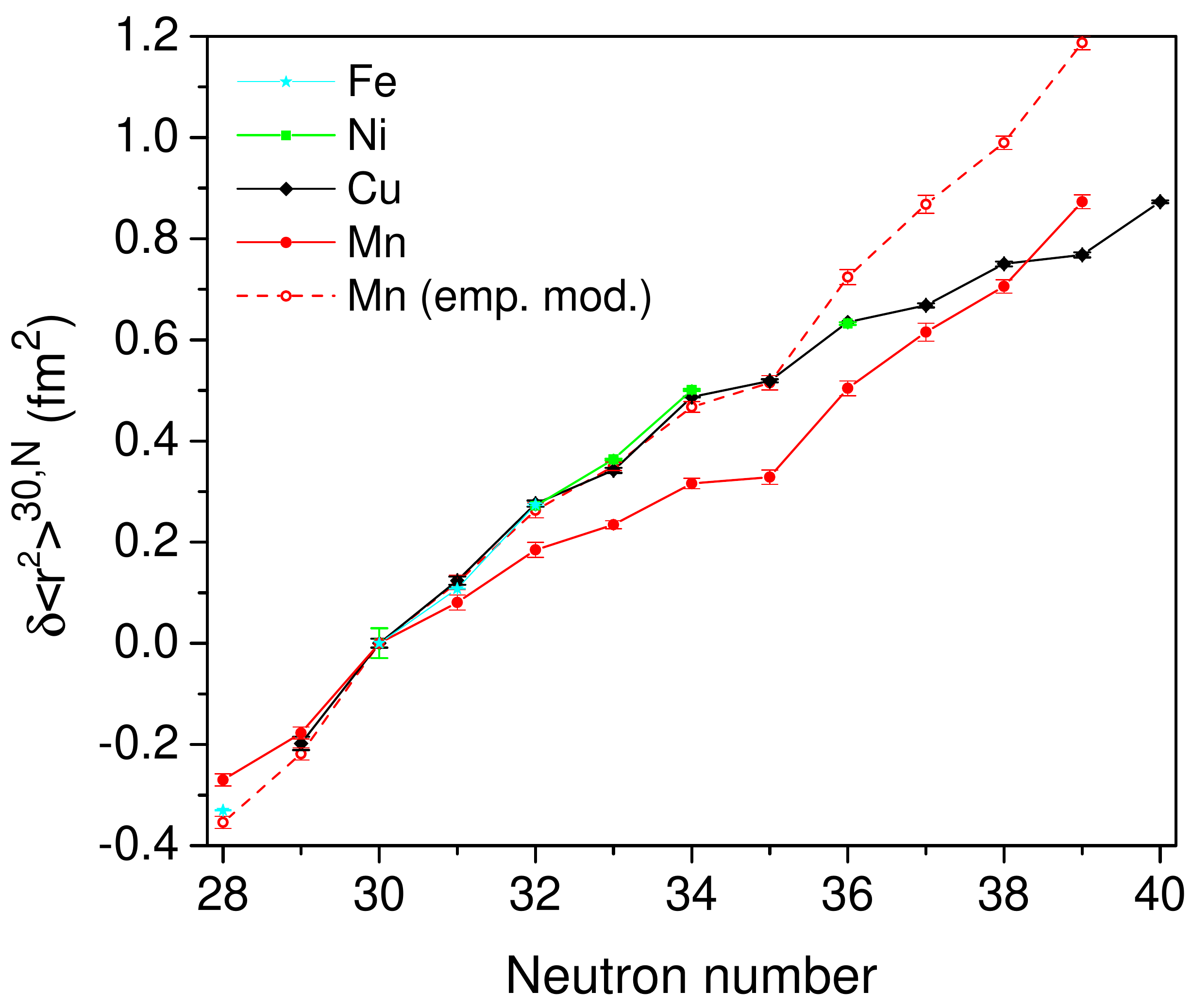}
	\caption{(Color online) Changes in mean-square charge radii of Mn, Fe, Ni and Cu with respect to $N=30$. The Mn radii obtained using the MCDF parameters are indicated by full symbols and full lines while the empirically modified ones are shown with open symbols and dashed lines. Only statistical errors are shown.}
	\label{Fig:Mn-Cu}
	\end{figure}
\\ A first observation is that the mean-square charge radii of the $4^+$ isomers in neutron-rich $^{58,60,62}$Mn are very close to the $1^+$ ground state radii, as indicated by the blue stars in Fig. \ref{Fig:chargeradii-systematics}. This suggests similar degrees of deformation and no sizable shape coexistence. Although the absolute difference in charge radii of the two states is small, the charge radius of the isomer in $^{58}$Mn is slightly smaller than that of the ground state, while it is slightly larger in $^{60}$Mn and $^{62}$Mn. Note that the difference in isomer and ground state charge radii is much larger on the neutron-deficient side. In the self-conjugate $N=Z=25$ nucleus $^{50}$Mn this can be understood in terms of the difference in isospin ($T=0$ and $T=1$), as was discussed previously for $^{38}_{19}$K$_{19}$ \cite{Bissell2014}. %For $^{52}$Mn no immediate explanation can be put forward.
\\  A strong effect of the $N=28$ shell closure is seen in all isotopic chains. Moreover, beyond $N=28$ the increase in charge radii is strikingly independent of $Z$, as pointed out in a study of the neutron-rich K isotopes \cite{Kreim2014}. Also the recently published radii of the Cu isotopes \cite{Bissell2016}, with 10 protons more than K, precisely follow this systematic trend. The slope of the Mn radii determined in this work on the other hand, appears to be slightly too low. This becomes especially evident in Fig.~\ref{Fig:Mn-Cu} which plots the changes in mean-square charge radii of $_{26}$Fe~\cite{Benton1997}, $_{28}$Ni~\cite{Steudel1980} and $_{29}$Cu \cite{Bissell2016} relative to $N=30$ together with the changes in mean-square charge radii of Mn determined in this work. Although nuclear structure effects appear as local features in $\delta \langle r^2 \rangle$, irrespective of the general slope, the similarities and differences with other isotopic chains become more clear when the Mn radii are empirically modified to reproduce the regional trend. This is achieved by increasing $M_\text{atom}$ by 5\% and keeping $F_\text{atom}$ constant. The corresponding $F_\text{ion}$ and $M_\text{ion}$ are calculated from the King plot results. This modification causes a tilt of the charge radii about the $^{55}$Mn reference point but keeps the local features unaltered as shown by the open symbols connected by a dashed line in Figs.~\ref{Fig:chargeradii-systematics} and \ref{Fig:Mn-Cu}. Note that although the charge radius at $N=35$ might seem anomalous at first, the use of the modified $M$ and $F$ shows that the apparent dip is rather due to a combination of odd-even staggering and a sudden increase in charge radii at $N=36$, as will be discussed next.
\begin{figure*}[hp] % for pagewide tables
\includegraphics[width=0.8\textwidth]{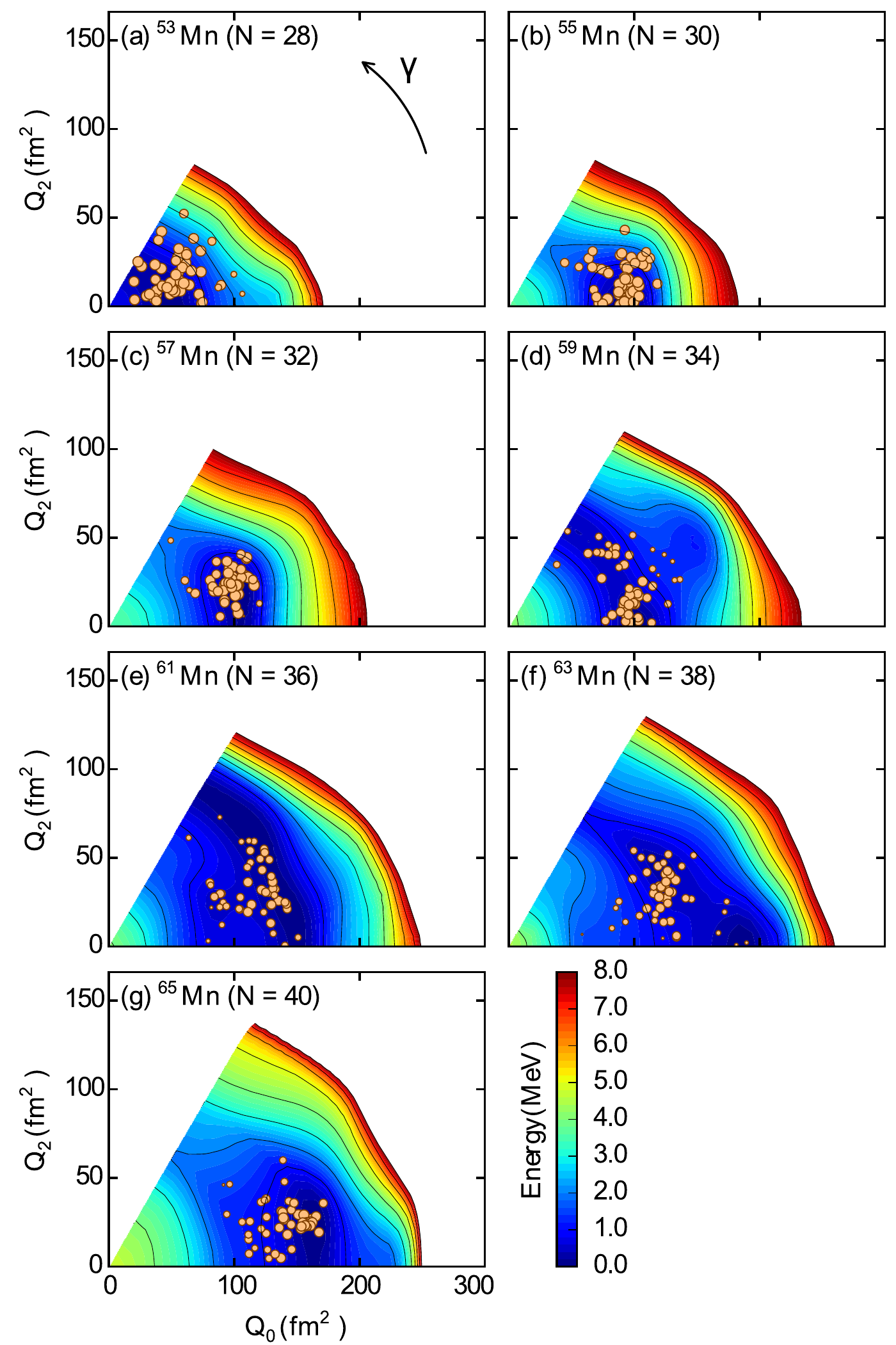}
    \caption{(Color online) Potential energy surfaces of odd-even $^{53-63}$Mn isotopes including the distribution of the MCSM basis states, depicted by the circles. The locations of the circles indicate the intrinsic shapes of the MCSM basis states and the sizes denote their importance in the total wave function, see reference \cite{Tsunoda2014} for more information. The energy scale is relative to the minimum of each potential energy surface, and the contour plots are shown up to a cut-off energy of 8 MeV.}
\label{Fig:MCSM}
\end{figure*}
\\ As shown in Fig.~\ref{Fig:Mn-Cu}, the modified Mn radii follow the Fe, Ni and Cu trend up to $N=35$ while beyond $N=35$ a clear divergence between the Mn and Cu isotopes is seen. Whereas in Cu the mean-square charge radii exhibit a weakly parabolic behavior between $N=28$ and $N=40$ \cite{Bissell2016}, in Mn a rapid increase in charge radii is observed from $N=36$ onwards. This observation is consistent with the increase in the spectroscopic quadrupole moments \cite{Babcock2016} and the behaviour of the magnetic moments \cite{Babcock2015, Heylen2015}, interpreted as arising from particle-hole excitations across $Z=28$ and $N=40$. From $N=36$ onwards, these magnetic and quadrupole moments can be described by large-scale shell model calculations only if excitations of neutrons from the $pf$ shell into the $g_{9/2}$ and $d_{5/2}$ orbits are included, as well as proton excitations across $Z=28$. %Since these orbitals belong to the next major shell (and in a simple harmonic oscillator potential the radii go like $N+3/2$ with $N$ the major shell), this results naturally in an increase in mean-square charge radius.
\\ A good reproduction of the experimental magnetic and quadrupole moments is achieved in the Monte Carlo Shell Model (MCSM) framework using the modified A3DA interaction \cite{Shimizu2012,Tsunoda2014} which includes the full $pf$ shell and $g_{9/2}d_{5/2}$ orbitals in the model space for both protons and neutrons \cite{Babcock2016}. This gives confidence in the predictive power of these calculations to discuss the shape of the Mn ground states. A constrained Hartree-Fock calculation using this A3DA shell model Hamiltonian provides the potential energy surface on which the distribution of the MCSM basis states (deformed slater determinants) as a function of the intrinsic quadrupole moments \mbox{$Q_{0} \propto \langle 2z^2 - x^2 - y^2 \rangle $} and \mbox{$Q_{2} \propto \langle x^2 - y^2 \rangle$} is plotted. Fig.~\ref{Fig:MCSM} presents these potential energy surfaces for the ground states of the odd-even $^{53-65}$Mn isotopes, showing the evolution of intrinsic shape between $N=28$ and $N=40$. At the $N=28$ shell closure the distribution of intrinsic shapes is centered around the spherical minimum. This changes towards moderately deformed prolate structures between $N=30$ and 34, then an additional increase in prolate deformation is seen in $N=36$ and 38 and a maximum is reached at $N=40$. However, a picture in which only axial symmetric deformation is considered too simple since from $N=34$ onwards a triaxial minimum appears and becomes dominant in $N=36, 38$ and $N =40$. Moreover, the observed spread in the distribution of intrinsic shapes points to considerable shape fluctuations in the ground states of the neutron-rich Mn isotopes. Such shape fluctuations were already suggested to be important in the low-energy structure of the neutron-rich $_{24}$Cr isotopes from a constrained Hartree-Fock-Bogoliubov plus local quasiparticle random-phase approximation study \cite{Sato2012}. 
\\ The sudden increase in experimental charge radii is in accordance with the predicted change in nuclear shape at $N=36$ although these calculations indicate that the change in the radii cannot be solely attributed to a change in static prolate deformation.%Hence, no $\beta_2$ deformation parameters are deduced from the observed radii.

\subsection{Charge radii and masses}
\begin{figure}[t]
	\includegraphics[width=\columnwidth]{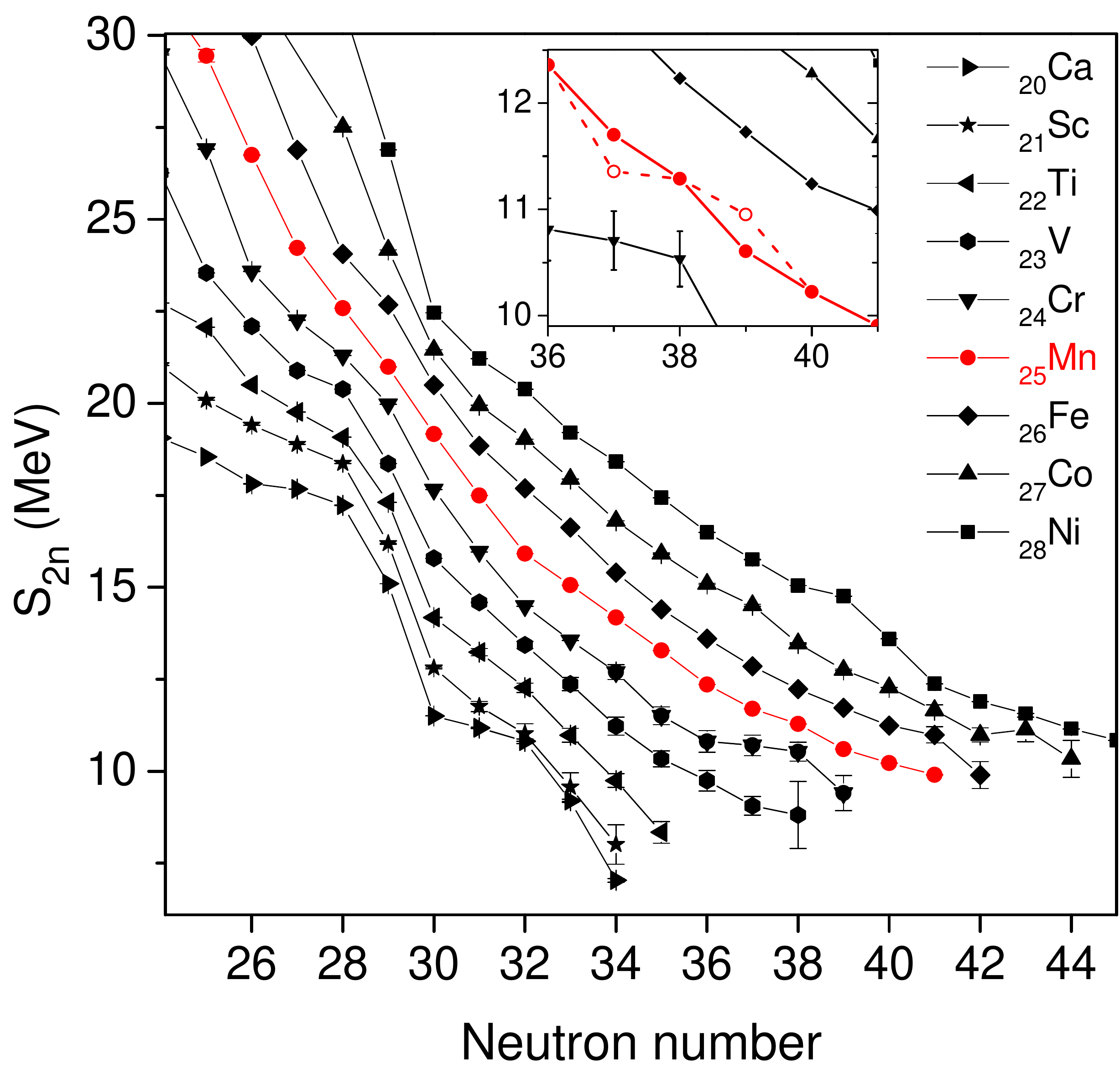}
	\caption{(Color online) Two-neutron separation energies plotted from $_{20}$Ca to $_{28}$Ni. Data taken from the {\sc{Ame2012}} atomic-mass evaluation \cite{AME2012} complemented with data from \cite{Wienholtz2013,Naimi2012}. The $^{62}$Mn binding energy is recalculated with the tentative Coulomb excitation information \cite{Gaffney2015} which influences the $S_{2n}$ at $^{62}$Mn and $^{64}$Mn. The inset shows the difference between the recalculated values (full line) and the original publication \cite{Naimi2012} (dashed line).}
	\label{Fig:S2n}
\end{figure}
In this section, the connection between the trend in mean-square charge radii and the masses of the Mn isotopes is examined. Across the nuclear chart, trends in charge radii are in general closely reflected in the course of the two-neutron separation energies, with clear examples in the $A\sim 100$ and rare-earth regions \cite{Charlwood2009,Cakirli2010}. However, this correspondence is not always straightforward, as shown in \cite{Charlwood2010} for the light mass region. For example, the prominent $N=28$ shell closure effect in the Mn charge radii is almost not visible in the two-neutron separation energies. Such discrepancies are not unexpected, since the charge radii and binding energies reflect different aspects of the nuclear wave function. In the context of shape coexistence for example, it becomes clear that states with very different quantum-mechanical structure can be almost degenerate in energy, thus a huge effect in nuclear size can be accompanied by a small change in nuclear binding.  In particular, for the Mn isotopes the $Z \approx N \approx 25$ (also known as the Wigner effect \cite{Vanisacker1995}) plays an important role in modifying the trend of the two-neutron separation energies close to $N = 28$. Furthermore, a change of ground-state structure when crossing $N=28$ might also bring an additional energy contribution to the trend close to the semi-magic $^{53}$Mn$_{28}$. Indeed, the spin changes from $5/2^-$ to $7/2^-$ and back to $5/2^-$ in $^{51,53,55}$Mn, along with a significant increase of their quadrupole moments going from $Q_s(^{53}$Mn)=+0.16(3) fm$^2$  at the shell closure to $Q_s=+0.42(7)$ fm$^2$ and $Q_s=+0.33(1)$ fm$^2$ in $^{51,55}$Mn, respectively \cite{Babcock2016,Charlwood2010}.
\\ Precise mass measurements of the neutron-rich Mn were presented by Naimi \emph{et al.}\ \cite{Naimi2012} noting that it was unclear whether the observed $4^+$ state in $^{62}$Mn was the ground or isomeric state, with a tentative assignment to the isomer. In the mean time, a Coulomb excitation experiment has been performed in which the $4^+$ state was proposed to lie 346 keV above the $1^+$ ground state \cite{Gaffney2015}. With this new information, the ground state \mbox{two-neutron} separation energies can be recalculated, as shown in Fig.~\ref{Fig:S2n}. The inset highlights the difference between the original (dashed line) and recalculated (solid line) values. While in the original publication a sudden increase in two-neutron separation energies at $N=38$ was observed using the  measured $^{62}$Mn mass, this effect is washed out using the recalculated ground state mass. Instead, the two-neutron separation energies gradually curve up towards $N=40$, which would rather be linked to a gradual onset of collectivity \cite{Fossion2002}. So although the onset of deformation seems to occur suddenly at $N=36$ in the charge radii, this is accompanied by a smooth change of the ground state binding energies.
\\ Similar to Mn, a smooth upwards curvature is also seen in the $_{26}$Fe isotopic chain while an abrupt discontinuity at $N=36$ is observed in the {\sc Ame2012} evaluation of $_{24}$Cr isotopes. However, new measurements show a much smoother behavior \cite{Meisel2016} although the reported errors are relatively large. High-precision mass measurements are required to clarify the issue. To complete the picture, mean-square charge radii measurements for the Cr and Fe isotopes would be highly desirable.

\subsection{\label{sec:theory}Duflo-Zuker analysis}
\begin{figure}[b]
	\includegraphics[width=\columnwidth]{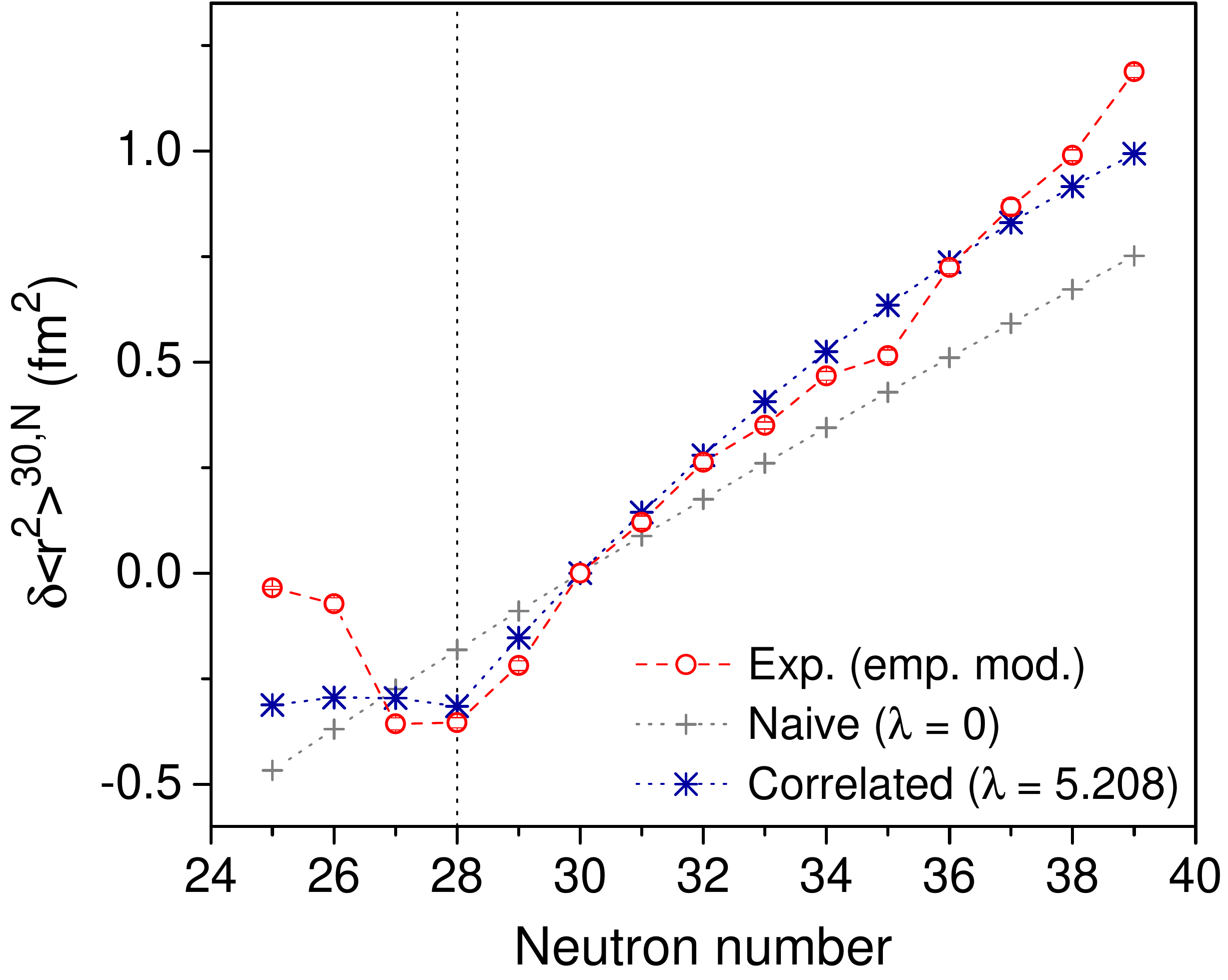}
	\caption{(Color online) Empirically modified changes in mean-square charge radii of Mn compared to the naive (eq.~\ref{eq:naive}) and correlated (eq.~\ref{eq:naive}+\ref{eq:correlated}) predictions using the Duflo-Zuker expression.}
	\label{Fig:Theory}
\end{figure}

To conclude the discussion, the changes in mean-square charge radii of the Mn isotopes are compared to theoretical calculations using the Duflo-Zuker  \cite{Duflo2002} formalism, shown in Fig.~\ref{Fig:Theory}. It is based on the fact that radii are single particle operators that depend on orbital occupancies and their sizes. A recent study \cite{Bonnard2016} has shown that it accounts for shell effects due to anomalously large orbits ($s_{1/2},\  p_{3/2}$ and $p_{1/2}$ in the neutron $sd$ and $pf$ shells, respectively), and leads to the first successful reproduction of the experimental changes in root mean-square charge radii in K and Ca.
\\ The Duflo-Zuker expression consists of two terms
 \begin{subequations}
\begin{align}
 \sqrt{\langle r_{\pi}^2\rangle} & =   A^{1/3}\left[\rho_0-\frac{\zeta}{2}\frac{t}{A^{4/3}}- \frac{\upsilon}{2}\left(\frac{t} {A}\right)^2\right]e^{(g/A)}\label{eq:naive} \\ 
& + \dfrac{\lambda}{A^{1/3}}\left[\frac{z(D_{\pi}-z)}{D_{\pi}^2}  \frac{n(D_{\nu}-n)}{D_{\nu}^2}\right] \label{eq:correlated}
\end{align}
\end{subequations}
where $t=N-Z$. Considering the terms \ref{eq:naive} and \ref{eq:correlated} separately, the first term (eq.~\ref{eq:naive}) is derived  by noting that $r_{\pi}$ is an isospin vector. Therefore, its square contains a scalar, a vector and a tensor, associated with coefficients $\rho_0,\,\zeta$ and $\upsilon$, respectively. The expression for neutrons, is obtained by reversing the sign of $t$, so the  difference in neutron and proton charge radius, the neutron skin, is given by 
\begin{align}
\Delta r_{\nu\pi}=\sqrt{\langle r_{\nu}^2\rangle} - \sqrt{\langle r_{\pi}^2\rangle}   = \frac{\zeta t e^{g/A}}{A}.
\end{align}
Here $e^{g/A}$ is a phenomenological correction factor which accounts for the larger radii in light nuclei. By a four-parameter fit to the known radii of isotopes with $A\leq 60$, $g=1.37$, $\rho_0 = 0.947$ fm, $\upsilon=0.295$ fm and $\zeta = 0.8$ fm are found giving a rms deviation of 0.04 fm. Using this first term as a naive approach which only depends on $t$ and $A$, a smooth variation of the charge radii is obtained as shown in Fig.~\ref{Fig:Theory}. The necessary shell effects are introduced by adding a second term (eq.~\ref{eq:correlated}) which depends on the orbital occupancies of the extruder-intruder (EI) valence spaces, delimited by $N,\, Z=6, 14, 28, 50, \ldots$~\cite[Sec.~IC]{rmp}. The degeneracy of each valence shell is then given by $D_{\pi,\nu}=8, 14, 22, \ldots$ and the number of active particles is indicated by $n, z$. For the Mn isotopes this becomes: $z = 11$ and $D_\pi = 14$ for the protons and $n = N - 14$, $D_\nu = 14$  for $N\leq28$ or $n = N - 28$ and $D_\nu = 22$ for $N>28$. Note that this second term becomes zero at the extruder-intruder shell closures.
\\ Once the second term (eq.~\ref{eq:correlated}) is included in the fit, slightly adapted fit parameters ($g=1.14$, $\rho_0 = 0.94$ fm, $\upsilon=0.334$ fm, $\zeta = 0.8$ fm and $\lambda$=5.208) are extracted and a good reproduction of the charge radii below $Z = 30$ is obtained. This reduces the rms deviation is to 0.02 fm. 
\\ The changes in mean-square charge radii obtained using the full Duflo-Zuker expression applied to the Mn isotopic chain (blue stars) are compared to the experimental values (red circles) in Fig.~\ref{Fig:Theory}. Between $N=28$ and $N=37$, the agreement is particularly good and similar to K and Ca, the strong increase in charge radii beyond $N=28$ is suggested to be associated with the large size of the (halo) neutron $p_{3/2}$ orbital \cite{Bonnard2016}. Due to isovector polarizability \cite{Bonnard2016}, an increase of neutron radii causes larger charge (proton) radii as well. Nevertheless, above $N=37$ and below $N=28$ there are clear deviations between theory and experiment. This indicates that the large deformations, shown to be important for the Mn isotopes in this and previous work \cite{Charlwood2010,Babcock2015,Heylen2015,Babcock2016}, are not fully taken into account in eq.~3. Since in the K and Ca isotopes no similar large onset of deformation below $N=28$ is observed, the mean-square charge radii are well reproduced \cite{Bonnard2016}.

\section{Conclusion}
The $^{51,53-64}$Mn isotopes have been studied in two collinear laser spectroscopy experiments at the COLLAPS beam line at ISOLDE, CERN. In this article, the isotope shifts and extracted mean-square charge radii were presented. 
\\ The electronic mass shift and field shift factors for the atomic $3d^5 4s^2 \ {}^6\text{S}_{5/2} \rightarrow 3d^5 4s 4p\ {}^6\text{P}_{3/2}$ and the ionic $3d^5 4s\ {}^5\text{S}_2 \rightarrow 3d^5 4p\ {}^5\text{P}_3$ transitions have been calculated in the multi-configurational Dirac-Fock framework. Because of the open $3d$ electron shell, these calculations are particularly complex and relatively large uncertainties on $M$ and $F$ are assumed. Alternatively, results are also presented in which $M$ is increased by 5\% in order to obtain a better agreement with the regional trend. Whether or not this modification is warranted will only become clear when more independently determined charge radii information in the region becomes available. It needs to be stressed once more that although the empirical modification causes a global tilt about the reference point, it does not affect the local irregularities in the charge radii. Since these local irregularities reveal the nuclear structure phenomena we are interested in, the physics discussion remains valid irrespective of the final choice of $F$ and $M$ (within the presented uncertainty).
\\ The nuclear structure along the Mn isotopic chain is discussed based on the available ground state properties. A clear connection between the moments and charge radii is observed, pointing to a good $N=28$ shell closure and a rapid onset of deformation towards $N=40$. The variation of the two-neutron separation energies on the other hand is rather smooth showing that these effects do not necessarily lead to discontinuities in the mass surface.
\\ Furthermore, the evolution of the intrinsic shapes in the Mn isotopic chain was investigated in the Monte Carlo Shell Model framework using the modified A3DA effective interaction. This suggests that the increase in charge radii from $N=35$ is not only due to a static increase in prolate deformation, but that also triaxiality and shape fluctuations are important.
\\ Finally, the changes in mean-square charge radii were discussed using the Duflo-Zuker model. The sharp increase in charge radii beyond $N=28$, common to all isotopes studied in the region so far, has been related to the occupancy of the neutron (halo) orbitals. However, below $N=28$ and above $N=37$ the experimental values depart from the theoretical predictions, which is explained as due to deformation.

\section*{Acknowledgements}

We would like to thank the ISOLDE technical group for their support and assistance. H. H.\ thanks V.\ Manea for the interesting discussions and the insightful suggestions related to the separation energies.
\\ This work was supported by the Belgian Research  Initiative on Exotic Nuclei (IAP-project P7/12), the FWO-Vlaanderen, GOA grant 10/010 and 10/015 from KU Leuven, the German Ministry for Education and Research (BMBF) under contract 05P15RDICA and 05P15SJCIA, the Max-Planck Society, the Science and Technology Facilities Council. This project has received funding through the European Union's Seventh Framework Programme for Research and Technological Development under Grant Agreements: 262010 (ENSAR), 267194 (COFUND), and 289191 (LA$^{3}$NET).

\bibliography{Mn,Al,Zuker}% Produces the bibliography via BibTeX.

\end{document}